\documentclass[aps,prb,showpacs,twocolumn,amsmath,amssymb,superscriptaddress]{revtex4}
\usepackage{graphicx}
\usepackage{dcolumn}
\usepackage{color}
\usepackage{bm}
\begin{document}

\title {Inelastic decay rate of quasiparticles in a two-dimensional spin-orbit coupled electron system}

\author{I.~A.~Nechaev}
\altaffiliation[Also at: ]{REC ``Physics and Chemistry of High-Energy Systems'', Tomsk State University,
634050, Tomsk, Russia.} \affiliation{Department of Theoretical Physics, Kostroma State University, 156961
Kostroma, Russia.} \affiliation{Donostia International Physics Center (DIPC), P. de Manuel Lardizabal, 4,
20018, San Sebasti{\'a}n, Basque Country, Spain}

\author{P.~M.~Echenique}
\author{E.~V.~Chulkov}
\affiliation{Donostia International Physics Center (DIPC), P. de Manuel Lardizabal, 4, 20018, San
Sebasti{\'a}n, Basque Country, Spain} \affiliation{Departamento de F{\'\i}sica de Materiales, Facultad de
Ciencias Qu{\'\i}micas, UPV/EHU and Centro Mixto CSIC-UPV/EHU, Apdo. 1072, 20080 San Sebasti\'an, Basque
Country, Spain}

\date{\today}

\begin{abstract}
We present a study of the inelastic decay rate of quasiparticles in a two-dimensional electron gas with
spin-orbit interaction. The study is done within the $G^0W^0$ approximation. The spin-orbit interaction is
taken in the most general form that includes both Rashba and Dresselhaus contributions linear in magnitude of
the electron 2D momentum. Spin-orbit interaction effect on the inelastic decay rate is examined at different
parameters characterizing the electron gas and the spin-orbit interaction strength in it.
\end{abstract}

\pacs{71.10.-w, 71.70.Ej}

\maketitle

\section{Introduction}

Nowadays, in condensed matter physics and semiconductor microelectronics, two-dimensional (2D) electron
system is one of the main objects of detailed study. Such a system is formed by, e.g., surface-state
electrons or electrons in semiconductor heterostructures. Phenomenon that is observed in such systems and
makes them of great interest, especially in context of spintronic applications, is spin-orbit interaction
(SOI). This interaction arise from the structure inversion asymmetry of potential confining the electron
system in directions perpendicular to the confinement plane (the Rashba spin-orbit
interaction\cite{Rashba_1960_1984}) and the bulk inversion asymmetry  that is present in semiconductor
heterostructures based on materials with a zinc-blende structure (the Dresselhaus spin-orbit
interaction\cite{Dresselhaus,DK_1986}). The Dresselhaus interaction depends on semiconductor material and
growth geometry, whereas the interaction strength of the Rashba SOI can be tuned via an externally applied
electric field perpendicular to the confinement plane.\cite{Studer_PRL2009} As a result, one can controllably
manipulate the spin in devices without recourse to an external magnetic field.\cite{Datta_1990,Nitta_1997}.

In order to efficiently exploit the mentioned phenomenon, a theoretical study of dynamics of electrons and
holes in the 2D spin-orbit coupled electron systems is needed. The most discussed and studied processes
concerning this problem are spin relaxation and spin dephasing.\cite{Zutic_2004} However, to our knowledge,
such crucial quasiparticle property as the lifetime caused by inelastic electron-electron scattering remains
still insufficiently studied. To all appearance the first attempt to analyze what effect the SOI has on the
quasiparticle lifetime has been made in Ref.~\onlinecite{Saraga_PRB2005}. In the work cited, a particular
case of the 2D electron gas (2DEG) with the Rashba SOI was considered at the limit of $E_R\ll E_F$, where
$E_F$ is the Fermi energy and $E_R=m^{\ast}\alpha^2/2$ with $\alpha$ and $m^{\ast}$ being the interaction
strength and the effective electron mass, respectively (unless stated otherwise, atomic units are used
throughout, i.e., $e^2=\hbar=m=1$.). Within the $G^0W^0$ approximation, it has been shown that in a small
vicinity of $E_F$ a modification of the lifetime due to the SOI is insignificant and does not depend on the
subband index of the spin-orbit split band. To go beyond the limits of Ref.~\onlinecite{Saraga_PRB2005}, in
Ref.~\onlinecite{NIA_CEV_FTT_2009} the inelastic lifetime (decay rate) of quasiparticles in the 2DEG with the
Rashba SOI has been studied within a wide energy region. For material parameters typical for
In$_x$Ga$_{1-x}$As 2DEGs, it has been revealed that modifications induced by the SOI and the dependence on
the subband index become noticeable, when the decay channel due to plasmon emission appears. The first joint
theoretical and experimental investigation of hole lifetimes in a 2D spin-orbit coupled electron system has
been done in Ref.~\onlinecite{Au111}. In addition to a demonstration of the weak influence of the SOI on hole
lifetimes by the case of the Au(111) surface state, a hypothetical system, where the SOI can have a profound
effect, has been considered.

In this work, we generalize the results on effect of the SOI on the quasiparticle lifetime. Within the
$G^0W^0$ approach with the screened interaction $W^0$ evaluated in the random phase approximation (RPA), we
study the inelastic decay rate of quasiparticles in a 2DEG with the Rashba and Dresselhaus interactions
linear in $k$---magnitude of the electron 2D momentum $\mathbf{k}$. In our $G^0W^0$-calculations, material
parameters suitable for InAs quantum wells are taken. We compare the inelastic decay rates calculated at
different ratios between the interaction strengths of the mentioned spin-orbit interactions. We show that on
the energy scale, for the taken material parameters, the main visible effect induced by the SOI is
modifications of the plasmon-emission decay channel via the extension of the Landau damping region. We also
consider a hypothetical small-density case, when in the 2D spin-orbit coupled electron system the Fermi level
is close to the band energy at $\mathbf{k}=0$. For such a system, we predict strong subband-index dependence
and anisotropy of the inelastic decay rate for electrons and appearance of a plasmon decay channel for holes.

\section{\label{sec:approximations}Approximations}

\begin{figure}[tbp]
\centering
 \includegraphics[angle=0,scale=1.0]{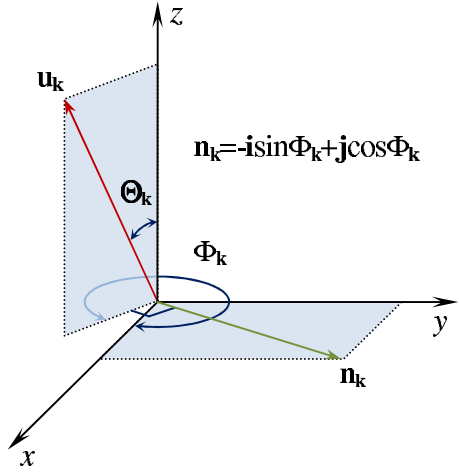}
\caption{(Color online) A spatial location of the vectors $\mathbf{u}_{\mathbf{k}}$ and
$\mathbf{n}_\mathbf{k}$ determining the spin-quantization axis with polar angles $\Theta_{\mathbf{k}}$ and
$\Phi_{\mathbf{k}}$ and the rotation axis, respectively.}\label{fig1}
\end{figure}

\begin{figure}[tbp]
\centering
 \includegraphics[angle=-90,scale=0.45]{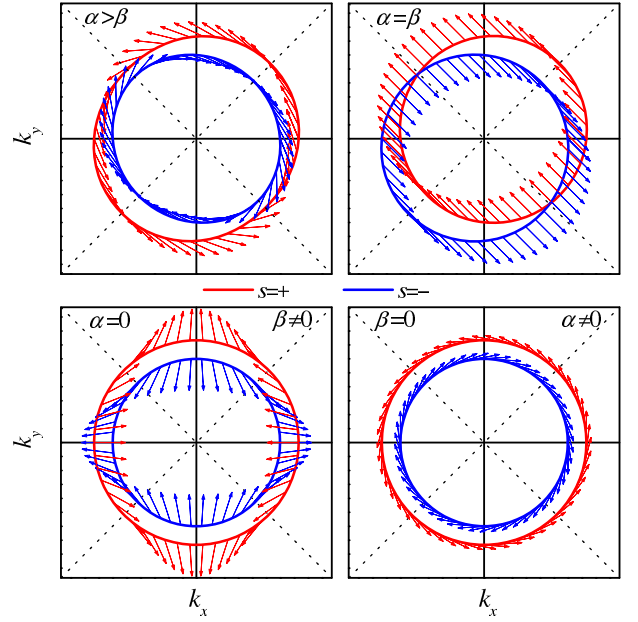}
\caption{(Color online) Fermi contours and spin orientations in the momentum plane for different values of
interaction strengths of the Rashba and Dresselhaus spin-orbit interactions.}\label{fig2}
\end{figure}

We consider a 2DEG described by the Hamiltonian $H=H_0+H_{SO}$ with $H_0=k^2/2m^{\ast}$ and the spin-orbit
contribution
\begin{equation}\label{Hamiltonian}
H_{SO}=\alpha\left(\sigma_{x}k_{y}-\sigma_{y}k_x\right)+ \beta\left(\sigma_{x}k_{x}-\sigma_{y}k_y\right)
\end{equation}
that includes both Rashba and Dresselhaus terms. The latter is written with the assumption that a quantum
well grown in [001] direction is considered. In Eq.~(\ref{Hamiltonian}), $k_{x,y}$ are the electron momenta
along the [100] and [010] cubic axes of the crystal, respectively, $\sigma_{x,y}$ are the Pauli matrices,
$m^{\ast}$ is the effective electron mass, $\alpha$ and $\beta$ are the interaction strengths for the Rashba
and Dresselhaus spin-orbit interactions. To bring the Hamiltonian to a diagonal form, we perform the rotation
in spin space generated by $U_{\mathbf{k}}=\exp[i(\bm{\sigma}\cdot
\mathbf{n}_{\mathbf{k}})\Theta_{\mathbf{k}}/2]$ dependent on the momentum $\mathbf{k}$. The rotation is
performed with the angle $\Theta_{\mathbf{k}}$ around the axis determined by $\mathbf{n}_{\mathbf{k}}$. A
positional relationship of the axis $\mathbf{n}_{\mathbf{k}}$ and the spin-quantization axis
$\mathbf{u}_{\mathbf{k}}$ is shown in Fig.~\ref{fig1}. We suppose that we deal with the in-plane spin
polarization, i.e., $\Theta_{\mathbf{k}}=\pi/2$. In the new, unitary transformed, spin basis the spin-orbit
contribution has the form\cite{PSH_remark}
\begin{eqnarray}\label{Hamiltonian_diagonal_form}
H'_{SO}&=&U^{\dag}_{\mathbf{k}}H_{SO}U_{\mathbf{k}} \nonumber\\
&=&-k\left[\alpha\sin(\varphi_{\mathbf{k}}-\Phi_{\mathbf{k}}) +
\beta\cos(\varphi_{\mathbf{k}}+\Phi_{\mathbf{k}})\right]\sigma_{z},
\end{eqnarray}
where the angle $\Phi_{\mathbf{k}}$ is related to the polar angle $\varphi_{\mathbf{k}}$ of the momentum
$\mathbf{k}$ as
\begin{equation}\label{Phi_k_def}
\tan\Phi_{\mathbf{k}}=-\frac{\alpha\cos\varphi_{\mathbf{k}}+\beta\sin\varphi_{\mathbf{k}}}
{\alpha\sin\varphi_{\mathbf{k}}+\beta\cos\varphi_{\mathbf{k}}}.
\end{equation}
Due to the diagonal form of $H'_{SO}$, the energy bands are simply given by\cite{GF_remark}
\begin{equation}\label{energies}
E_{\mathbf{k}s}=\frac{\mathbf{k}^2}{2m^{\ast}}+sk\left[\alpha\sin(\varphi_{\mathbf{k}}-\Phi_{\mathbf{k}}) +
\beta\cos(\varphi_{\mathbf{k}}+\Phi_{\mathbf{k}})\right]
\end{equation}
and correspond to the wave functions $\psi'_{\mathbf{k}s}(\mathbf{r})=e^{i\mathbf{k}\mathbf{r}}|-s\rangle$
with the subband index $s=\pm(\downarrow,\uparrow)$, where $\downarrow,\,\uparrow$ are the spin components in
the new spin basis. This means that for the initial, untransformed, Hamiltonian we have the following
eigenstates $\psi_{\mathbf{k}s}(\mathbf{r})=U_{\mathbf{k}}\psi'_{\mathbf{k}s}(\mathbf{r})$. The spin
orientation in $\mathbf{k}$ space reads as (see Fig.~\ref{fig2})
\begin{equation}\label{averaged_spin}
\langle\psi_{\mathbf{k}s}|\bm{\sigma}|\psi_{\mathbf{k}s}\rangle=s
\left(\begin{array}{c}
    \cos\Phi_{\mathbf{k}} \\
    \sin\Phi_{\mathbf{k}} \\
    0\\
\end{array}\right)
\end{equation}
Note that the case with $\alpha\neq0$ and $\beta=0$ (pure Rashba) is characterized by the angle
$\Phi_{\mathbf{k}}=\varphi_{\mathbf{k}}-\pi/2$, whereas in the situation with $\beta\neq0$ and $\alpha=0$
(pure Dresselhaus) one has $\Phi_{\mathbf{k}}=2\pi-\varphi_{\mathbf{k}}$. In the special case of
$\alpha=\beta$, the angle $\Phi_{\mathbf{k}}=-\pi/4$ does not depend on $\varphi_{\mathbf{k}}$.

The inelastic decay rate (inverse lifetime $\tau^{-1}_s(\mathbf{k})$ caused by inelastic electron-electron
scattering) is determined by the imaginary part of the matrix elements of the quasiparticle self-energy
$\langle\Sigma_s(\mathbf{k},\omega)\rangle=\langle\psi_{\mathbf{k}s}(\mathbf{r}_1)|
\Sigma(\mathbf{r}_1,\mathbf{r}_2;\omega)|\psi_{\mathbf{k}s}(\mathbf{r}_2)\rangle_{\mathbf{r}_1\mathbf{r}_2}$
at the energy $\omega=E_{\mathbf{k}s}$ as
$\Gamma_s(\mathbf{k})=2|\mathrm{Im}\langle\Sigma_s(\mathbf{k},E_{\mathbf{k}s})\rangle|$. At the Hartree-Fock
(HF) mean-field level, these elements are totally real and have the form
\begin{equation}\label{HF_sigma}
\langle\Sigma^{HF}_s(\mathbf{k})\rangle=-\sum_{s'}
\int\frac{d\mathbf{q}}{(2\pi)^2}F^{ss'}_{\mathbf{k},\mathbf{q}}f_{\mathbf{q}s'}v_c(\mathbf{k}-\mathbf{q}),
\end{equation}
where $v_c(\mathbf{k})=2\pi/(|\mathbf{k}|\varepsilon_0)$ is the bare Coulomb interaction with $\varepsilon_0$
being the static dielectric constant. The factors $F^{ss'}_{\mathbf{k},\mathbf{p}}
=\left[1+ss'\mathbf{u}_{\mathbf{k}}\cdot\mathbf{u}_{\mathbf{p}}\right]/2$ come from $|\langle
s'|U^{\dag}_{\mathbf{p}}U_{\mathbf{k}}|s\rangle|^2$ and $f_{\mathbf{k}s}$ is the Fermi factor. Such a form
(\ref{HF_sigma}) is similar to the exchange contribution to the single-particle energies considered in
Refs.~\onlinecite{Chen_PRB1999, Chesi_PRB2007, Juri_PRB2008} in the pure Rashba case. However, in order to
examine quasiparticle lifetimes, one has to go beyond the HF approximation. The simplest variant is the
$G^0W^0$ approximation (for details about the approximation, we refer the reader to Refs.
\onlinecite{NIA_CEV_FTT_2009} and \onlinecite{Aluminium_2008}). Within such an approximation, we arrive at
the following expression for the imaginary part of the mentioned matrix elements
\begin{eqnarray}\label{ME_SIGMA_IM_below_EF}
\mathrm{Im}\langle\Sigma_s(\mathbf{k},\omega)\rangle&=&-\sum_{s'}
\int\frac{d\mathbf{q}}{(2\pi)^2}F^{ss'}_{\mathbf{k},\mathbf{q}}f_{\mathbf{q}s'}\\
&\times&\mathrm{Im}W^0(\mathbf{k}-\mathbf{q},\omega-E_{\mathbf{q}s'})\theta(E_{\mathbf{q}s'}-\omega),\nonumber
\end{eqnarray}
when $\omega<E_F$, and
\begin{eqnarray}\label{ME_SIGMA_IM_above_EF}
\mathrm{Im}\langle\Sigma_s(\mathbf{k},\omega)\rangle&=&\sum_{s'}
\int\frac{d\mathbf{q}}{(2\pi)^2}F^{ss'}_{\mathbf{k},\mathbf{q}} [1-f_{\mathbf{q}s'}]\\
&\times&\mathrm{Im}W^0(\mathbf{k}-\mathbf{q},\omega-E_{\mathbf{q}s'})\theta(\omega-E_{\mathbf{q}s'}),\nonumber
\end{eqnarray}
when $\omega>E_F$. In these equations, $\theta(x)$ is the step function and the screened interaction
\begin{equation}\label{W_RPA}
W^0(\mathbf{q},\omega)=v_c(\mathbf{q})\left[1-P^0(\mathbf{q},\omega)v_c(\mathbf{q})\right]^{-1}
\end{equation}
is defined by the RPA irreducible polarizability (see also Refs.~\onlinecite{Pletyukhov_PRB2006} and
\onlinecite{Badalyan_PRB2009}, where the retarded part of $P^0$ was examined)
\begin{eqnarray}\label{IP0_mom}
P^0(\mathbf{q},\omega)&=&\sum_{ss'}\int\frac{d\mathbf{k}}{(2\pi)^2}F^{ss'}_{\mathbf{k},\mathbf{k}+\mathbf{q}}\nonumber\\
&\times&\left\{\frac{(1-f_{\mathbf{k}+\mathbf{q}s})f_{\mathbf{k}s'}}{\omega+E_{\mathbf{k}s'}-E_{\mathbf{k}+\mathbf{q}s}+i\eta}\right.\\
&-&\left.\frac{f_{\mathbf{k}+\mathbf{q}s}(1-f_{\mathbf{k}s'})}
{\omega+E_{\mathbf{k}s'}-E_{\mathbf{k}+\mathbf{q}s}-i\eta}\right\}.\nonumber
\end{eqnarray}

\section{\label{sec:results_discussion}Results and discussion}

Using the example of an InAs quantum well, we take the effective mass $m^{\ast}=0.023$ (see, e.g.,
Ref.~\onlinecite{Knap_PRB1996}) and the static dielectric constant $\varepsilon_0=14.55$ (see, e.g.,
Ref.~\onlinecite{Lorimor_1965}). The interaction strength of the Dresselhaus SOI is chosen to be
$\beta=1.6\times10^{-11}$ eV\,m to simulate a quite narrow quantum well.\cite{Knap_PRB1996} At the ratio
$\alpha/\beta=2.4$ (see, e.g., Ref.~\onlinecite{Giglberger_PRB2007}), we have the Rashba interaction strength
$\alpha=3.8\times10^{-11}$ eV\,m. The electron density is put at $n_{2D}=2.55\times10^{11}$ cm$^{-2}$ that
corresponds to the Fermi energy $E_F\approx26$ meV.

\begin{figure}[tbp]
\centering
 \includegraphics[angle=-90,scale=0.4]{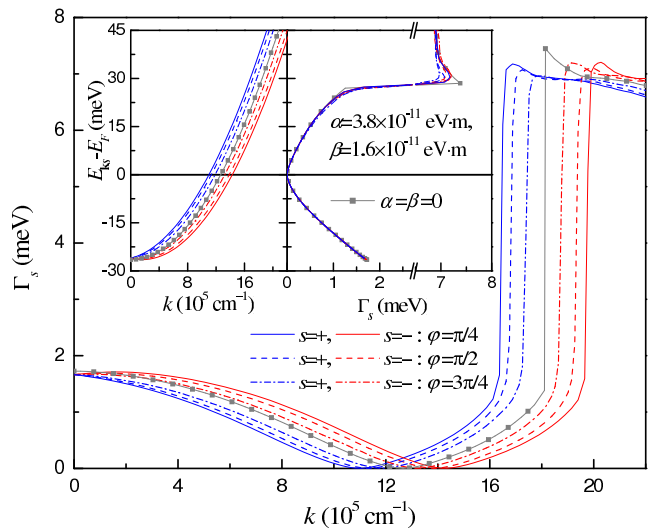}
\caption{(Color online) Inelastic decay rate $\Gamma_s$ as a function of $k$ at several values of the polar
angle $\varphi_{\mathbf{k}}$ in the case of the ratio $\alpha/\beta=2.4$. Inset: The corresponding energy
bands $E_{\mathbf{k}s}$ (at the left) and the same $\Gamma_s$ as a function of $E_{\mathbf{k}s}$ measured
from the Fermi energy (at the right). Also, for reference, the case of the 2DEG without the SOI
($\alpha=\beta=0$) is presented.}\label{fig3}
\end{figure}

Fig.~\ref{fig3} shows our results on the inelastic decay rate $\Gamma_s(\mathbf{k})$ obtained with the
material parameters listed above. Two main points caused by the spin-orbit splitting of the band make the
considered 2DEG different from that without the SOI. These are an angle-dependent relative shift of
$\Gamma_+$ and $\Gamma_-$ on the momentum scale and some smoothing of sharp forms of the peak caused by
opening of the plasmon decay channel. The former reflects the fact that subbands of the split band reach the
same energy at different momenta, while the latter originates from the extension of the Landau damping region
(the region where plasmons decay into single-particle excitations\cite{Bruus_Flensberg}) due to appearance of
inter-subband transitions (for a detailed discussion of the screening properties of the 2DEG with the SOI we
refer the reader to Refs.~\onlinecite{Pletyukhov_PRB2006} and \onlinecite{Badalyan_PRB2009}). This extension
varying with the polar angle $\varphi_{\mathbf{k}}$ leads to a nonzero plasmon linewidth, when the plasmon
spectrum enters into the SOI-induced damping region.

In order to show what effect the SOI has on the inelastic decay rate for different subbands, in the inset of
Fig.~\ref{fig3}, by setting up a correspondence between $\Gamma_s(\mathbf{k})$ and $E_{\mathbf{k}s}$ via the
momentum $\mathbf{k}$, we plot the decay rate as a function of energy. On first glance, it may seem that we
have an ordinary energy dependence of the decay rate as in a 2DEG without the SOI: the quadratic behavior
with the logarithmic enhancement in the vicinity of the Fermi energy with $\Gamma_s=0$ at $E_F$ and the jump
above the Fermi energy, which is caused by opening the plasmon decay channel for excited
electrons.\cite{GV_electron_liquid} However, on examining the energy dependence of $\Gamma_s$ in detail, we
can say that due to the finite plasmon linewidth the plasmon decay channel manifests itself at lower
energies, when it occurs in a 2DEG without the SOI. The same reason leads to reduction in the jump. Also, we
can reveal distinctions between $\Gamma_+$ and $\Gamma_-$, which become noticeable, when the plasmon-emission
decay channel appears, and increase upon moving from $\varphi_{\mathbf{k}}=\pi/4$ to $3\pi/4$. An analysis of
the inelastic mean-free path (IMFP) $\lambda_s(\mathbf{k})=|\mathbf{\nabla}_{\mathbf{k}}E_{\mathbf{k}s}|
/\Gamma_s(\mathbf{k})$ as a function of energy has shown that, as a consequence of the distinctions between
$\Gamma_+$ and $\Gamma_-$, the IMFP of electrons can vary with the subband index. For example, at
$\varphi_{\mathbf{k}}=3\pi/4$ for electrons this variation can reach, e.g., $\sim 8$\%.

The obtained results can be understood by inspecting constant-energy contours shown in Fig.~\ref{fig2} with
mental drawing of possible transitions selected by the factors $F^{ss'}_{\mathbf{k},\mathbf{q}}$ of
Eqs.~(\ref{ME_SIGMA_IM_below_EF}) and (\ref{ME_SIGMA_IM_above_EF}) at a given $\varphi_{\mathbf{k}}$ (see
also Ref.~\onlinecite{Au111}). Actually, for each subband ($s=\pm$) one has a set of intra- and inter-subband
transition momenta as arguments of $\mathrm{Im}W^0$. For the chosen material parameters and for
$\varphi_{\mathbf{k}}=\pi/4$, these momenta do not vary considerably with the subband index $s$. For
$\varphi_{\mathbf{k}}=3\pi/4$ differences in both intra- and inter-subband transitions for $s=+$ and $s=-$
become already sensible for values of $\mathrm{Im}W^0$, especially in the vicinity of plasmon peaks of the
latter.

\begin{figure}[tbp]
\centering
 \includegraphics[angle=-90,scale=0.4]{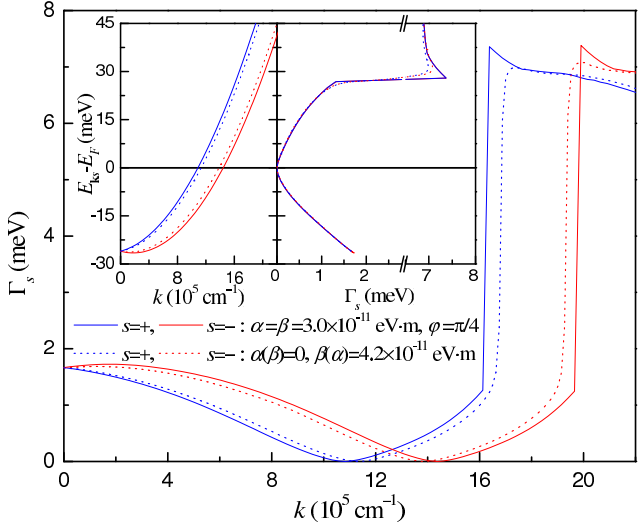}
\caption{(Color online) Same as in Fig.~\ref{fig3}, but for the ratio $\alpha/\beta=1$. Also, the pure Rashba
and Dresselhaus cases are presented.}\label{fig4}
\end{figure}

Now, remaining $m^{\ast}$, $\varepsilon_0$, $n_{2D}$, and $E_F$ unchanged, we consider the case of
$\alpha/\beta=1$ [$\alpha=\beta=3.0\times10^{-11}$ eV\,m] and the pure Dresselhaus (Rashba) case
[$\beta(\alpha)=4.2\times10^{-11}$ eV\,m and $\alpha(\beta)=0$]. The case of equal interaction strengths,
when the Rashba and Dresselhaus interactions can cancel each other, is distinguished by various significant
effects reported in the literature (see, e.g., Refs.~\onlinecite{Averkiev_PRB1999, Schliemann_PRL2003,
Bernevig_PRL2006, Koralek_Nature_2009}). In this case, one has the 2D electron system with two uncoupled spin
components (see Fig.~\ref{fig2}), each of which demonstrates the properties peculiar to a 2DEG without the
SOI.\cite{Badalyan_PRB2009} Our results\cite{remark_isotropic} on the inelastic decay rate at
$\alpha/\beta=1$ are shown in Fig.~\ref{fig4}. The sharp edges of the plasmon contribution are evidence of
the fact that there is no modifications of the Landau damping region induced by the SOI. As is seen from the
inset of the figure, due to the shifting property $E_{\mathbf{k}+}=E_{\mathbf{Q}+\mathbf{k}-}$, where
$Q_{x}=Q_{y}=\sqrt{8}m^{\ast}\alpha$, the $\Gamma_+$ and $\Gamma_-$ curves coincide and have the form of that
in a 2DEG without the SOI.

In the pure Rashba or pure Dresselhaus cases (see Fig.~\ref{fig4}), the resulting $\Gamma_s$ does not tell
the difference between spin orientations in the momentum plane, which correspond to the Rashba or Dresselhaus
SOI (see Fig.~\ref{fig2}). As well as before, we have the relative shift (but angle-independent) on the
momentum scale and main modifications induced by the SOI in the energy region, where a quasiparticle can
decay into plasmons.

\begin{figure}[tbp]
\centering
 \includegraphics[angle=-90,scale=0.4]{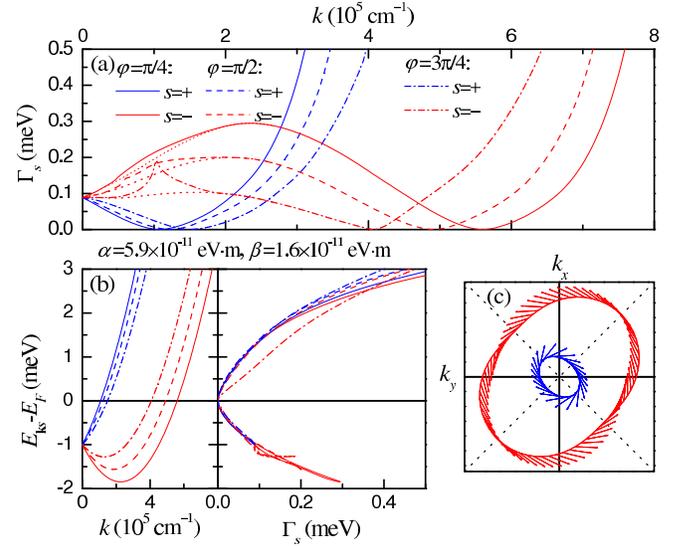}
\caption{(Color online) (a) Inelastic decay rate $\Gamma_s$ as a function of $k$ at several values of the
polar angle $\varphi_{\mathbf{k}}$ in the small electron density case at $\alpha/\beta=3.7$. Dotted lines
represent $\Gamma_s$ without the plasmon contribution (see the text). (b) The corresponding energy bands
$E_{\mathbf{k}s}$ (at the left) and the same $\Gamma_s$ as a function of $E_{\mathbf{k}s}$ measured from the
Fermi energy (at the right). (c) Spin orientations in the momentum plane for the considered electron
gas.}\label{fig5}
\end{figure}

All the considered cases meet the condition of $E_{RnD}\ll E_F$, where
$E_{RnD}=m^{\ast}(|\alpha|+|\beta|)^2/2$ is the measure of influence of the SOI on the band structure.
However, as is partly discussed in Ref.~\onlinecite{Au111}, in two-dimensional electron systems with much
greater $E_{RnD}$ as compared to $E_F$ the inelastic decay rate can substantially differ from that in the
2DEG without the SOI. A striking example of such a system is that formed by surface-state electrons in
ordered surface alloys,\cite{Surf_Alloys} which are very promising materials for spintronics applications. In
order to predict how the inelastic decay rate can behave in a system, where $E_{RnD}\sim E_F$, we consider
the hypothetical case with the unchanged $m^{\ast}=0.023$, $\varepsilon_0=14.55$, and
$\beta=1.6\times10^{-11}$ eV\,m, but with $\alpha=5.9\times10^{-11}$ eV\,m ($\alpha/\beta=3.7$) and
$n_{2D}=2.04\times10^{10}$ cm$^{-2}$, which give $E_F=1.0$ meV and $E_{RnD}=0.85$ meV. The obtained results
are presented in Fig.~\ref{fig5}.

The main feature we would like to note first is that for holes the decay rate $\Gamma_-$ as a function of $k$
has an ``outgrowth'' at $k<k_{min}$, where $k_{min}\equiv m^{\ast} [\alpha^2 + \beta^2 + 2\alpha\beta
\sin(2\varphi_{\mathbf{k}})]^{1/2}$ --- the momentum, at which $E_{\mathbf{k}-}$ has a minimum. A close
analysis of the imaginary part of the screened interaction $W^0$ and the region of integration in
Eq.~(\ref{ME_SIGMA_IM_below_EF}) has shown that the outgrowth is caused by opening of the plasmon decay
channel for transitions between the $s=-$ and $s'=+$ subbands. It is important that in a 2DEG without the SOI
such a channel is impossible for holes.\cite{GV_electron_liquid}

As is evident from the figure, in this case on the energy scale we have a strong anisotropy of the inelastic
decay rate. Also the presented curves clearly demonstrate that the latter depends strongly on the subband
index $s$ of the spin-orbit split band. Keeping in mind that the index $s$ distinguishes spin components, we
can say that the subband-index dependence reflects a spin asymmetry of the inelastic decay rate
$\Gamma_-/\Gamma_+$ for a given direction. The asymmetry shows  its worth most brightly in the
$\varphi_{\mathbf{k}}=3\pi/4$ direction. In fact, in that very direction there are significant distinctions
in $\Gamma_-$ and $\Gamma_+$ as functions of the exciting energy and, as a consequence, in the corresponding
IMFP for electrons. For instance, the ratio $\Gamma_-/\Gamma_+$ is about 3 at $\sim 0.5$ meV and about 2 at
$\sim 1.0$ meV. At further increasing of energy, the ratio continues to decrease.

Note that the IMFP spin asymmetry makes a basis of the spin filter effect observed in hot electron transport
through a ferromagnetic (see, e.g., Refs.~\onlinecite{Zutic_2004} and \onlinecite{UFN_2009}). In the
considered case of the quantum well, the spin asymmetry is not such big as in ferromagnetics (see, e.g.,
Ref.~\onlinecite{NIA_CEV_FTT_FECO_2009}), but, as distinct from the latters, values of the IMFP spin symmetry
depend strongly on direction and can be tuned by external electric field.

\section{\label{sec:conclusions}Conclusions}

In conclusion, we have presented a study of the inelastic decay rate of quasiparticles in a two-dimensional
electron gas with the $k$-linear spin-orbit interaction that includes both Rashba (interaction strength
$\alpha$) and Dresselhaus (interaction strength $\beta$) contributions. In this study, the electron gas is
characterized by material parameters suitable for [001]-grown InAs quantum wells. We have considered the
cases of $\alpha>\beta>0$, $\alpha=\beta$, $\alpha=0$ ($\beta>0$), and $\beta=0$ ($\alpha>0$). The cases meet
the condition of $E_{RnD}\ll E_F$, where $E_{RnD}=m^{\ast}(|\alpha|+|\beta|)^2/2$ is the measure of influence
of the spin-orbit interaction on the band structure.

As compared to a two-dimensional electron gas without the spin-orbit interaction, we have revealed a relative
shift of the inelastic decay rates for different subbands of the spin-orbit split band on the momentum scale.
Also, except for the case of equal interaction strengths, we have found a some smoothing of sharp forms of
the peak concerned with opening of the plasmon decay channel for electrons. We have shown that, on the energy
scale, in this very region distinctions between the decay rates for different subbands become noticeable.
These distinctions depend on the polar angle $\varphi_{\mathbf{k}}$ and cause the inelastic mean free path to
be angle- and subband-dependent. As to the case of $\alpha=\beta$, due to the shifting property, the decay
rate as a function of energy has the form of that in a two-dimensional electron gas without the spin-orbit
interaction.

In order to predict how the inelastic decay rate can behave in a system, where $E_{RnD}\sim E_F$, we have
considered the hypothetical case of small electron density. We have revealed that in such a system the decay
rate demonstrates strong anisotropy and subband dependence within all the considered interval of momenta and
exciting energies. Since the subband dependence can be interpreted as a spin asymmetry of the decay rate in a
given direction of $\mathbf{k}$, one can expect the spin-filter effect driven by externally applied electric
field. Also, we have found that in the system with $E_{RnD}\sim E_F$ holes can decay into plasmons, what is
impossible in a two-dimensional electron gas without the spin-orbit interaction.

\section*{\label{sec:acknowledgments}Acknowledgments}
We acknowledge partial support from the University of the Basque Country (Grant No. GIC07IT36607) and the
Spanish Ministerio de Ciencia y Tecnolog\'{i}a (Grant No. FIS2007-66711-C02-01). Calculations were partly
performed on SKIF-Cyberia supercomputer of Tomsk State University.



\begin{thebibliography}{00}
\bibitem{GF_remark} Note that Eq.~(\ref{energies}) can be cast into the form that is frequently used in the literature. Actually,
from
$\left(E_{\mathbf{k}s}-\mathbf{k}^2/2m^{\ast}\right)^2=\left(k\left[\alpha\sin(\varphi_{\mathbf{k}}-\Phi_{\mathbf{k}})
+ \beta\cos(\varphi_{\mathbf{k}}+\Phi_{\mathbf{k}})\right]\right)^2$ with the help of Eq.~(\ref{Phi_k_def})
we arrive at $\left(E_{\mathbf{k}s}-\mathbf{k}^2/2m^{\ast}\right)^2=
k^2[\alpha^2+\beta^2+2\alpha\beta\sin(2\varphi_{\mathbf{k}})]$. The latter can be solved as
$E_{\mathbf{k}s}=\mathbf{k}^2/2m^{\ast}+s k[\alpha^2+\beta^2+2\alpha\beta\sin(2\varphi_{\mathbf{k}})]^{1/2}$.
However, in such an expression the subband index $s=\pm$ distinguishes the inner- and outer-branch and, e.g.,
in the case of $\alpha=\beta$ does not correspond to spin components.
\bibitem {remark_isotropic} We have chosen the direction $\varphi_{\mathbf{k}}=\pi/4$, in which we can reproduce the momentum
dependence of the inelastic decay rate peculiar for a 2DEG without the SOI. For each spin components, such a
dependence becomes isotropic if one shifts the origin in the momentum plane on the vector $s\mathbf{Q}/2$.
\end{thebibliography}
\end{document}